\documentclass[aps,longbibliography,showpacs,twocolumn,superscriptaddress,amsmath,amssymb,verbatim]{revtex4-1}

\usepackage[thinlines]{easytable}
\usepackage{graphicx}% Include figure files
\usepackage{lmodern}
\usepackage{epstopdf}
\usepackage{dcolumn}% Align table columns on decimal point
\usepackage{bm}% bold math
\usepackage{subfigure}
\usepackage{makecell}
\usepackage{amsmath} %math
\usepackage[pdftex,colorlinks=true,citecolor=blue,linkcolor=blue,urlcolor=blue,bookmarks=true]{hyperref}

\usepackage{color}
\usepackage{textcomp}
\definecolor{red}{rgb}{1,0,0}

\definecolor{blue}{rgb}{0,0,1}

\definecolor{green}{rgb}{0,1,0}

\begin{document}
	\preprint{APS}

\title{Magnetism and spin dynamics of an $S$ = 3/2 frustrated trillium lattice  antiferromagnet K$_{2}$CrTi(PO$_{4}$)$_{3}$
 }

\author{J. Khatua}
\affiliation{Department of Physics, Indian Institute of Technology Madras, Chennai 600036, India}
\affiliation{Department of Physics, Sungkyunkwan University, Suwon 16419, Republic of Korea}
%\affiliation{Department of Physics, Indian Institute of Technology Madras, Chennai 600036, India}
\author{Suheon Lee}
\affiliation{Center for Artificial Low Dimensional Electronic Systems, Institute for Basic Science, Pohang 37673, Republic of Korea}
\author{Gyungbin Ban}
\affiliation{Department of Physics, Sungkyunkwan University, Suwon 16419, Republic of Korea}
\author{Marc Uhlarz}
\affiliation{Dresden High Magnetic Field Laboratory (HLD-EMFL), Helmholtz-Zentrum Dresden-Rossendorf, 01328 Dresden, Germany}
\author{Kwang-Yong Choi}
\email{choisky99@skku.edu}
\affiliation{Department of Physics, Sungkyunkwan University, Suwon 16419, Republic of Korea}
\author{P. Khuntia}
\email[]{pkhuntia@iitm.ac.in}
\affiliation{Department of Physics, Indian Institute of Technology Madras, Chennai 600036, India}
\affiliation{Quantum Centre of Excellence for Diamond and Emergent Materials, Indian Institute of Technology Madras,
	Chennai 600036, India.}

\date{\today}

\begin{abstract}
	Competing magnetic interactions, frustration-driven quantum fluctuations, and spin-correlation offer an ideal route for the experimental  realization of  emergent quantum phenomena and exotic quasi-particle excitations in three-dimensional frustrated magnets. In this context, trillium lattice, wherein magnetic ions decorate a three-dimensional chiral network of corner-shared equilateral triangular motifs, provides a viable ground.
	Herein, we present the crystal structure, magnetic susceptibility,  specific heat, electron spin-resonance (ESR), muon spin-relaxation ($\mu$SR)  results on the polycrystalline samples of  K$_{2}$CrTi(PO$_{4}$)$_{3}$ wherein the Cr$^{3+}$ ions form a perfect trillium lattice without any detectable anti-site disorder.  The Curie-Weiss fit of the magnetic susceptibility data above 100 K yields a Curie-Weiss temperature $\theta_{\rm CW}$ =  $-$ 23 K, which indicates the presence of dominant antiferromagnetic interactions between $S$ = 3/2 moments of Cr$^{3+}$ ions. The specific heat measurements reveal the occurrence of two consecutive phase transitions, at temperatures $T_{\rm L}$ = 4.3 K and $T_{\rm H}$ = 8 K, corresponding to two different magnetic phases. Additionally, it unveils the existence of short-range spin correlations above the ordering temperature $T_{\rm H}$. The power-law behavior of ESR linewidth suggests the persistence of  short-range spin correlations   over a relatively wide critical region $(T-T_{\rm H}$)/$T_{\rm H}>$ 0.25 in agreement with the specific heat results. The $\mu$SR results provide concrete evidence of two different phases corresponding to two transitions, coupled with the critical slowing down of  spin fluctuations above $T_{\rm L}$ and persistent spin dynamics below $T_{\rm L}$, consistent with the thermodynamic results.  
	Moreover, the $\mu$SR results reveal the coexistence of static and dynamic local magnetic fields below $T_{\rm L}$, signifying the presence of complex magnetic phases owing to the entwining of spin correlations and competing magnetic interactions in this three-dimensional frustrated magnet.	

\end{abstract}
\maketitle
\section{Introduction}
Frustrated quantum materials, where competing interactions between localized spin moments and frustration-induced strong quantum fluctuations prevent classical N\'eel order, are highlighted as promising contenders for the discovery of emergent physical phenomena such as quantum spin liquid (QSL), and spin ice with exotic excitations that goes beyond conventional symmetry breaking paradigms in condensed matter \cite{Balents2010,PhysRevLett.17.1133,KHUNTIA2019165435,KHATUA20231}.\\ A QSL state is a highly entangled state, in which frustration-induced strong quantum fluctuations  defy long-range magnetic order even at  absolute zero temperatures, despite strong exchange coupling between magnetic moments in the host spin-lattice \cite{RevModPhys.89.025003,Wen2019,Savary_2016,doi:10.1126/science.aay0668,ANDERSON1973153}. Instead, two spins in the corresponding lattice form resonating spin-singlet pairs or long-range entangled states with exotic fractionalized excitations such as spinons coupled to emergent gauge fields \cite{PhysRevB.44.2664}. Experimental realization of  fractional quantum numbers, their identification and interaction are of profound importance in elucidating the underlying mechanism of some of the groundbreaking phenomena such as high-temperature superconductivity, fractional Hall effect, topological spin-textures, and monopoles in quantum condensed matter \cite{BASKARAN1987973,anderson1987resonating,PhysRevLett.127.097002,RevModPhys.87.457,doi:10.1146/annurev-conmatphys-020911-125058,RevModPhys.80.1083,Phillips2020,GOBEL20211}  that  could have far-reaching ramifications in both fundamental physics and innovative quantum technologies \cite{RevModPhys.80.1083}. The realization of QSL state is well established in one-dimensional magnets, however, its identification in higher dimensional spin system remains a long-standing challenge in modern condensed matter physics.\\
In the quest of this long-sought goal, significant effort has been devoted to the exploration of two-dimensional (2D) frustrated lattices \cite{KHATUA20231}, which has led to the  identification of a few QSL candidates in 2D systems \cite{jeon2023oneninth,Arh2022}. Meanwhile, three-dimensional (3D) frustrated spin-lattices such as hyperkagome and pyrochlore lattice are found to host QSL as well. Despite the fact that quantum fluctuations are less pronounced in 3D lattices, certain frustrated 3D lattices, including  transition metal-based hyperkagome lattices PbCuTe$_{2}$O$_{6}$ \cite{PhysRevLett.116.107203,Chillal2020} and Na$_{4}$Ir$_{3}$O$_{8}$ \cite{PhysRevLett.115.047201,PhysRevLett.113.247601} along with  pyrochlore lattice NaCaNi$_{2}$F$_{7}$ \cite{Plumb2019,PhysRevB.92.014406,Cai_2018}  as well as rare-earth-based pyrochlore lattice Ce$_{2}$Zr$_{2}$O$_{7}$ \cite{Gao2019}, shows signatures of QSL state driven by frustration-induced strong quantum fluctuations \cite{PhysRevB.106.104404}. 
However, the experimental realization of QSL in 3D spin lattice remains scarce, as most of the candidate materials show spin-freezing or magnetic ordering owing to perturbing terms in the spin Hamiltonian, lattice imperfections, and exchange anisotropy. In this respect,  it is pertinent to investigate  promising frustrated magnets on 3D spin lattices wherein competition between emergent degrees of freedom offers an alternate route to stabilize  exotic quantum and topological states \cite{KHATUA20231,Broholmeaay0668,Kivelson2023,doi:10.1080/00107514.2023.2284522,Khatua2022}
\\
To date, hexagonal-based layered frustrated magnets have been studied rigorously  for their 2D QSL and certain cubic lattice systems for their 3D counterparts \cite{Nguyen2021}. Recently, there is growing interest in non-centrosymmetric frustrated quantum materials which crystallize in the
cubic space group $P2_{1}3$ \cite{bradley2010mathematical,Kakihana2017,PhysRevB.82.014410}. The lack of inversion symmetry  in such non-centrosymmetric  materials gives rise to interesting physical phenomena owing to anisotropic Dzyaloshinskii-Moriya  (DM) interactions \cite{PhysRevB.88.214402,bradley2010mathematical,dzyaloshinskii1964theory,Ding2021,Ramakrishnan2019,izyumov1984modulated}. For example, non-centrosymmetric cubic
quantum materials are known to host a wide range of
quantum phenomena including  magnetic skyrmions in insulator Cu$_{2}$OSe$_{2}$O$_{3}$ \cite{doi:10.1126/sciadv.aat7323}, and in itinerant magnets MnSi \cite{doi:10.1126/science.1166767}, and  FeGe \cite{Yu2011},  and  topological Hall effect in MnGe \cite{PhysRevLett.106.156603}.
Apart from quantum phenomena induced by DM interactions, the spin frustration induced quantum phenomena in such 3D non-centrosymmetric chiral magnets have garnered significant attention in the quest to achieve distinct quantum phenomena. It is observed that in such quantum materials that crystallizes in the cubic structure ($P2_{1}3$)  the magnetic ions form a  trillium lattice i.e., a 3D chiral network of corner-shared equilateral triangular motifs  which provides the origin of spin frustration and may host a myriads of frustration-induced physical phenomena such as QSL. It is a well known that a geometrically frustrated lattice has the potential to harbor frustration-induced new quantum phenomena, as observed in intermetallic compounds. Notable examples include the presence of a  dynamic state above the transition temperature in EuPt(Si/Ge) \cite{PhysRevB.104.045145}, a spin-ice-like state in CeIrSi \cite{PhysRevB.100.134442}, and the occurrence of pressure-induced quantum phase transitions in MnSi \cite{PhysRevB.55.8330}.\\
 Nevertheless, oxide-based materials having a single trillium lattice structure are extremely rare. Recently, the trillium lattice, composed of corner-sharing equilateral triangles with six nearest neighbors, has emerged as an alternative route to further explore the influence of quantum fluctuations in the 3D QSL state and to realize unique quantum phenomena \cite{PhysRevB.74.224441,PhysRevB.82.014410}.
Remarkably, a recent study on the langbeinite family member K$_{2}$Ni$_{2}$(SO$_{4}$)$_{3}$, which crystallizes in  a cubic ($P$2$_{1}$3) crystal structure, reveals that Ni$^{2+}$ ($S$ = 1) ions form a coupled trillium lattice \cite{PhysRevLett.127.157204}. This material demonstrates several
magnetic sublattices  with competing magnetic interactions
 and exhibits a 3D QSL state induced by an applied magnetic field \cite{PhysRevLett.127.157204}. More recently, continuum spin excitations driven by strong quantum fluctuations are revealed by inelastic neutron scattering measurement on single crystals of K$_{2}$Ni$_{2}$(SO$_{4}$)$_{3}$ \cite{PhysRevLett.131.146701}. Furthermore, despite possessing a high spin
moment, the signature of a spin-liquid state has been observed in the trillium compounds   KSrFe$_{2}$(PO$_{4}$)$_{3}$ (Fe$^{3+}$; $S$ = 5/2) \cite{10.1063/5.0096942} and Na[Mn(HCOO)$_{3}$] (Mn$^{2+}$; $S$ = 5/2) \cite{PhysRevLett.128.177201}, that invokes  further investigation of the ground state and associated quasi-particle excitations in  analogous trillium lattice antiferromagnets. In addition, 
the spin correlations, interplay between competing
interactions and the topology of the electronic band structure in non-centrosymmetric magnets stabilizing in 3D spin lattice
make them an exciting class of materials with potential technological applications \cite{doi:10.1146/annurev-matsci-080921-110002,PhysRevMaterials.7.114402}.\\
Herein, we focus on a new phosphate langbeinite K$_{2}$CrTi(PO$_{4}$)$_{3}$ (henceforth KCTPO), wherein Cr$^{3+}$ ions with $S$ = 3/2 form a trillium lattice without any detectable anti-site disorder between constituent atoms.  Magnetization data reveals the presence of dominant antiferromagnetic interactions between $S$ = 3/2 moments of  Cr$^{3+}$ ions. Specific heat measurements demonstrate two magnetic transitions, namely at $T_{\rm L}$ = 4.3 K and $T_{\rm H}$ = 8 K, as well as the development of short-range spin correlations above $T_{\rm H}$, which is supported by critical-like behavior of the ESR line width.  Furthermore, $\mu$SR data uncovers an intriguing evolution of static and dynamic spin correlations, notably below $T_{\rm L}$, and  a critical slowing down of spin dynamics above $T_{\rm L}$. Our experimental  findings reveal a complex magnetic landscape with coexisting states, successive phase transitions, and dynamic behavior of Cr$^{3+}$ moments  decorating a frustrated 3D trillium spin-lattice over a wide temperature regime.

\begin{figure*}
 	\centering
 	\includegraphics[width=\textwidth]{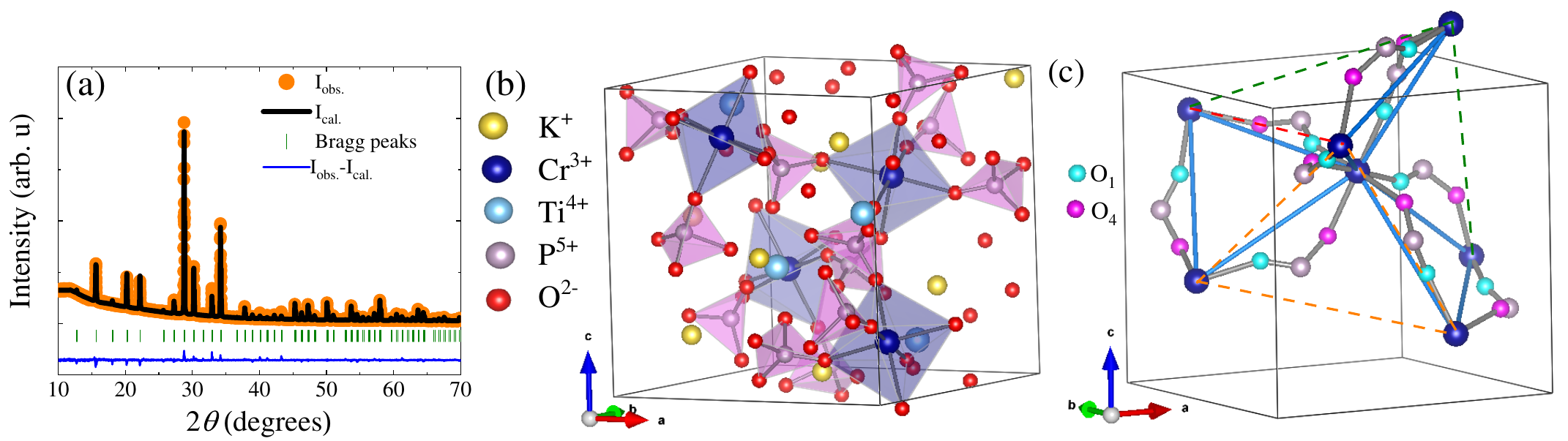}
 	\caption{(a) Rietveld refinement pattern of the room temperature powder x-ray diffraction data of K$_{2}$CrTi(PO$_{4}$)$_{3}$. The orange circle, black line, olive vertical bars, and blue line show the experimentally observed points, the result of Rietveld fitting, expected Bragg reflection positions, and the difference between observed and calculated intensities, respectively. (b) Schematic depicting the one unit cell of K$_{2}$CrTi(PO$_{4}$)$_{3}$.
 	The Cr$^{3+}$ ions form distorted CrO$_{6}$ octahedra (blue), and the P$^{5+}$ ions form PO${_{4}}$ tetrahedra (pink). These are connected through a shared oxygen ion, resulting in the formation of a Cr-O-P-O-Cr super-exchange bridge with nearest CrO$_{6}$ octahedra.  (c) The nearest-neighbor Cr$^{3+}$ ions (solid blue line; 6.10 \AA) arrange themselves in a trillium lattice, featuring six nearest neighbors. Trillium lattice is composed of three  motifs of equilateral triangle of Cr$^{3+}$ ions with possible nearest-neighbor superexchange routes. The second (8.32 \AA), third (9.79 \AA), and fourth (10.06 \AA) nearest neighbors are shown by the dashed orange, red, and olive lines, respectively. 
 	Other atoms are omitted to ensure a clearer view of the trillium lattice and potential nearest-neighbor exchange paths. }{\label{KCTPO1}}.
 \end{figure*}
\section{Experimental details}  Polycrystalline samples of KCTPO were prepared by a conventional solid-state  reaction  method. The appropriate stoichiometric amounts of K$_{2}$CO$_{3}$ (Alfa Aesar, 99.997 \text{\%}), Cr$_{2}$O$_{3}$ (Alfa Aesar, 99.97 \text{\%}), TiO$_{2}$ (Alfa Aesar, 99.995 \text{\%})  and (NH$_{4}$)$_{2}$HPO$_{4}$ (Alfa Aesar, 98 \text{\%})  were mixed. Prior to use, the reagent K$_{2}$CO$_{3}$ was preheated in air at 100$^{\circ}$C to eliminate moisture.
The stoichiometric mixture was pelletized, and the pellet was sintered at 300$^{\circ}$C for 4 hrs. This sintering procedure was repeated at several intermediate temperatures before annealing the sample at 800$^{\circ}$C for 48 hrs to achieve a single-phase composition. 
Powder x-ray diffraction (XRD)
data were collected using a smartLAB Rigaku x-ray diffractometer with Cu K$_{\alpha}$ radiation ($\lambda $ = 1.54 {\AA}) at room temperature.\\ Magnetization measurements were performed using
the VSM option of Physical Properties Measurement System (PPMS, Quantum Design)  in the temperature range of 2 K $\leq$ \textit{T} $\leq$ 300 K and in magnetic fields up to 7 T. Specific heat measurements were performed  using PPMS by thermal relaxation method in the temperature range of 1.9 K $\leq$ \textit{T} $\leq$ 250 K and in magnetic fields up to 7 T. High-field magnetization measurement was conducted at Dresden High Magnetic Field Laboratory, sweeping a magnetic field up to 30 T at 4 K using a nondestructive pulsed magnet. The obtained data were scaled to the isothermal magnetization taken with PPMS at 4 K.\\
 X-band ($\nu$ =  9.5 GHz) electron  spin  resonance  (ESR)  measurements were performed  using  a Bruker EMXplus-9.5/12/P/L  spectrometer with  a continuous He flow cryostat in the temperature range of 4 K $\leq$ $T$ $\leq$ 300 K. The $\mu^{+}$SR experiments in zero field and in weak transverse magnetic field (32 G) were performed at the surface
muon beamline M20  at TRIUMF in the temperature range 1.93 K $\leq$ $T$ $\leq$ 80 K. Powder samples (approximately 0.5 g) were placed into a thin envelope composed of Mylar tape coated with aluminum ($\sim$ 50 $\mu$m thick), which was mounted on a Cu fork sample stick.
A standard $^4$He flow cryostat was employed to achieve the base temperature of 1.93 K for TRIUMF/M20. The obtained $\mu$SR data was analyzed using the musrfit software package \cite{SUTER201269}.
%\begin{figure}
%	\centering
%\includegraphics[width=0.5\textwidth]{bondleng}
%\caption{(a) Trillium lattice is composed up of three  motifs of equilateral triangle of Cr$^{3+}$ ions with possible nearest-neighbor superexchange routes. The second (8.32 \AA) and third (9.79 \AA) nearest neighbors are shown by the dashed blue and orange lines, respectively. (b) A magnified depiction of a single triangular motif and its related bond lengths and angles with surrounding non-magnetic atoms is outlined in Table 3. }{\label{angles}}.
%\end{figure}
\section{results}
\subsection{Rietveld refinement  and crystal structure } 
In order to ensure phase purity and determine atomic parameters of  the polycrystalline samples of KCTPO, the Rietveld refinement of powder XRD data was performed using GSAS software \cite{Toby:hw0089}. For the Rietveld refinement, the initial atomic parameters were obtained from references \cite{ISASI2000303,Boudjada:a15338,Norberg:os0095}. The Rietveld refinement reveals that KCTPO crystallizes in a cubic space group ($P2_{1}3$) and the resulting refinement pattern is shown in Fig.~\ref{KCTPO1} (a). No secondary phase was found, confirming the successful synthesis of KCTPO in its single phase. The observed sharp and well-defined XRD peaks imply that high-quality polycrystalline samples have been used in this study.  The obtained lattice parameters as well as goodness factors  from the Rietveld refinement of powder XRD are tabulated in table \ref{table}. Our analysis shows the absence of any detectable anti-site disorders between the atoms that constitute KCTPO.\\ Figure~\ref{KCTPO1} (b) depicts the one unit cell of KCTPO, where Cr$^{3+}$ ions form distorted  CrO$_{6}$ octahedra with nearest-neighbor O$^{2-}$ that are separated from adjacent CrO$_{6}$ octahedra by PO$_{4}$ tetrahedra.  The bonding between the CrO$_{6}$ octahedra and TiO$_{4}$ tetrahedra through a common oxygen ion is expected to construct a 3D Cr-O-P-O-Cr superexchange pathway between two nearest-neighbor Cr$^{3+}$ ions (see table~\ref{tableangle}).  Most interestingly,  the nearest-neighbor Cr$^{3+}$ ions (6.10 \AA) form a trillium lattice, i.e., a 3D network of corner-shared equilateral
triangular lattice with  coordination number six (Fig.~\ref{KCTPO1} (c)), potentially serving as a platform for hosting a frustrated spin-lattice in 3D. \\
From a structural point of view,  KCTPO belongs to the langbeinite family K$_{2}$$M_{2}$(SO$_{4}$)$_{3}$ ($M$ = Mg, Co, Ni,..), which crystallizes in the cubic crystal structure (space structure $P$2$_{1}$3). Two crystallographic 4$a$ sites occupy divalent magnetic ions that form a double trillium lattice in the sulfate langbeinites without any anti-site disorder \cite{PhysRevLett.127.157204}.  Among various phosphate langbeinites, the recently reported phosphate langbeinite KSrFe$_{2}$(PO$_{4}$)$_{3}$ shows a similar configuration, featuring a double trillium lattice with two magnetic sites of Fe$^{3+}$ ions and K/Sr anti-site disorder \cite{10.1063/5.0096942}. Unlike the aforementioned double trillium lattice-based langbeinites, KCTPO, a phosphate langbeinite, accommodates a single magnetic site (4$a$) hosting Cr$^{3+}$ ions, alongside another 4$a$ site occupied by non-magnetic Ti$^{4+}$ ions. This characteristic positions KCTPO as a considerably simpler 3D lattice, suited for the experimental realization of theoretically proposed physical phenomena intrinsic to the trillium lattice \cite{PhysRevB.78.014404,Kakihana2017}.
 \begin{figure*}
	\centering
	\includegraphics[width=\textwidth]{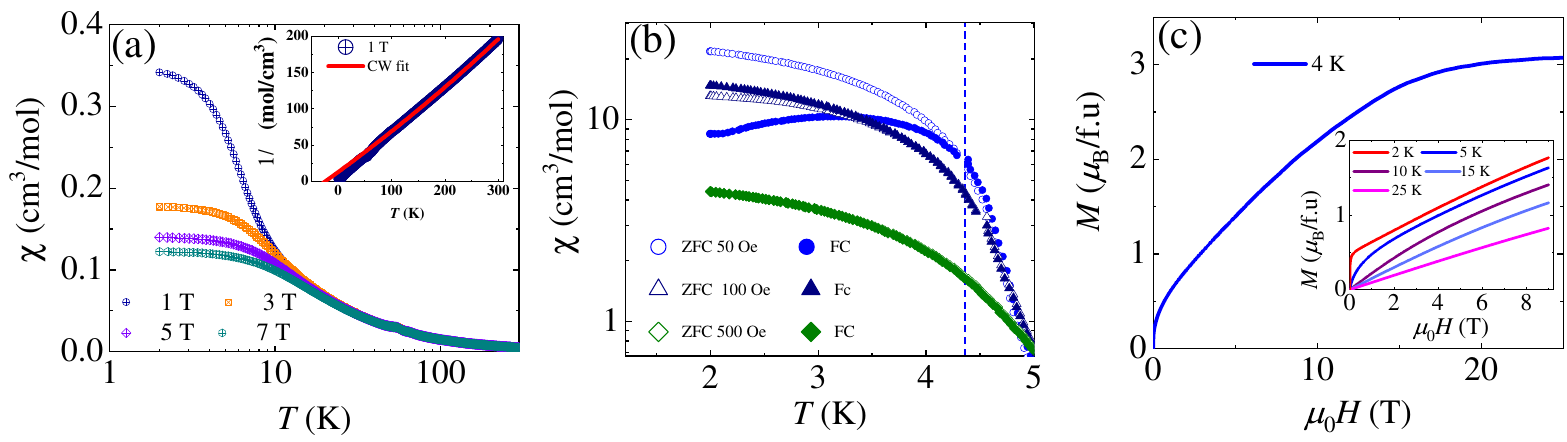}
	\caption{(a) Temperature dependence of magnetic susceptibility, $\chi(T)$, of K$_{2}$CrTi(PO$_{4}$)$_{3}$ in several magnetic fields. The inset depicts the temperature dependence of inverse magnetic susceptibility at $\mu_{0}H$ = 1 T. The red line represents  the Curie-Weiss fits to the  high-temperature inverse susceptibility data.   (b) Temperature dependence of zero-field cooled (ZFC) and field-cooled (FC)  magnetic susceptibility in the temperature range 2 K $\leq$ $T$ $\leq$ 5 K in magnetic fields  0.005 T $\leq$ $\mu_{0}H$ $\leq$ 0.05 T.  The dotted vertical line is at temperature $T_{\rm L}$ = 4.3 K, below which  ZFC and FC bifurcation begins in a magnetic field of $\mu_{0}H$ = 0.005 T.   (c) Magnetization as a function of an external magnetic field up to 30 T at 4 K. The inset shows the isotherm magnetization up to 9 T at several temperatures.  }{\label{KCTPO2}}.
\end{figure*}
\begin{table}
	\caption{\label{table}  Rietveld refinement of x-ray diffraction data at 300 K yielded structural parameters of K$_{2}$CrTi(PO$_{4}$)$_{3}$. (Space group: $P$2$_{1}$3, $a = b= c$ = 9.796 \AA , $\alpha = \beta = \gamma = 90^{\circ}$
		and $\chi^{2}$ = 2.86, R$_{\rm wp}$ = 5.29 \text{\%}, R$_{\rm p}$ = 3.56 \text{\%}, and R$ _{\rm exp}$ = 1.84\text{\%})}
	\begin{tabular}{c c c c c  c c} % <-- Alignments: l for left, c for center, and r for right, with vertical lines in between
		\hline \hline
		Atom & Wyckoff position & \textit{x} & \textit{y} &\textit{ z}& Occ.\\
		\hline 
		Cr & 4$a$ & 0.666 \ \ & 0.666 \ \ & 0.666 \ \ & 1 \\
		K$_{1}$ & 4$a$ & 0.960 \ \ & 0.960\ \ & 0.960 \  & 1 \\
		K$_{2}$ & 4$a$ & 0.186\ \ & 0.186\ \ & 0.186 \ \ & 1 \\
		Ti & 4$a$ & 0.391\ \ & 0.391\ \ & 0.391\ \ & 1 \\
		P & 12$b$ & 0.474\ \ & 0.703\ \ & 0.368 \ \ & 1 \\
		O$_{1}$ & 12$b$ & 0.609\ \ & 0.860\ \ & 0.737\ \ & 1 \\
		O$_{2}$ & 12$b$ & 0.463\ \ & 0.314\ \ & 0.232\ \ & 1 \\
		O$_{3}$ &  12$b$ & 0.506\ \ & 0.573\ \ & 0.313\ \ & 1 \\
		O$_{4}$ & 12$b$ & 0.585\ \ & 0.721\ \ & 0.457\ \ & 1 \\
		\\	
		\hline
	\end{tabular}
	%\end{ruledtabular}
\end{table}
\begin{table}
	\caption{\label{tableangle}  Bond lengths and angles between atoms that result in distinct antiferromagnetic interactions between Cr$^{3+}$ spins. }
	\begin{tabular}{c c c c c  c c}
		\hline \hline
		Bond length (\AA)  & Bond angle ($^\circ$) &\\
		\hline 
		Cr-O$_{1}$ =  2.101 & $\angle$ Cr-O$_{1}$-P = 153.510 & \\
		Cr-O$_{4}$ = 2.262 & $\angle$  O$_{1}$-P-O$_{4}$ = 120.635  & \\
		P-O$_{4}$ = 1.405 & $\angle$ P-O$_{4}$-Cr = 143.804 & \\ 
		P-O$_{1}$ = 1.273 &   $\angle$Cr-Cr-Cr = 60\\
		Cr-P = 3.494 &$\angle$ Cr-P-Cr = 128.355\\
		O$_{1}$-O$_{4}$ = 2.32 & $\angle$ Cr-O$_{1}$ -Cr = 154.692\\       
		\\	
		\hline
	\end{tabular}
	%\end{ruledtabular}
\end{table}
\subsection{Magnetic susceptibility}
Figure~\ref{KCTPO2} (a) depicts the temperature dependence of magnetic susceptibility $\chi(T)$ in several magnetic fields up to  7 T. 
Upon lowering temperature, a sharp
increase of $\chi(T)$ was observed below 10 K, tending to saturate in a magnetic field $\mu_0H$ = 1 T. This behavior indicates the presence of long-range magnetic order in KCTPO consistent with the specific heat result presented in this work. However,  this behavior is suppressed significantly in magnetic fields $\mu_{0}H \geq 3$ T. A minor bump around 50 K that appeared in all magnetic fields is attributed to residual oxygen trapped in the polycrystalline sample that was wrapped with teflon  during measurement \cite{PhysRevB.100.064423}. This signal is not an intrinsic characteristic of the investigated compound in this study. \\
In order to determine the dominant magnetic interaction between the $S$ = 3/2 moments of Cr$^{3+}$ ions, the linear region (100 K $\leq$ $T$ $\leq$ 300 K ) of the 1/$\chi(T)$ data in a magnetic field $\mu_{0}H$ = 1 T was fitted by the Curie-Weiss (CW) law, i.e., $\chi = \chi_{0}$ + $C$/($T$ $-$ $\theta_{\rm CW}$). Here, $\chi_{0}$ is the sum of temperature-independent core diamagnetic susceptibility ($\chi_{\rm core}$) and Van Vleck paramagnetic susceptibility ($\chi_{\rm VV}$), $C$ is the Curie constant used to estimate the effective magnetic moment ($\mu_{\rm eff}$ = $\sqrt{8C}$ $\mu_{\rm B}$) and  $\theta_{\rm CW}$ is the Curie-Weiss temperature associated with the exchange interaction between magnetic  moments of Cr$^{3+}$ ions. The corresponding  CW fit (red line in the inset of Fig.~\ref{KCTPO2} (a)) yields $\chi_{0}$ = $-$7.42 $\times$ 10$^{-4}$ cm$^{3}$/mol, $C$ = 1.88 cm$^{3}$ K/mol, and $\theta_{\rm CW}$ = $-$23 $\pm$ 0.15 K. The estimated effective magnetic moment $\mu_{\rm eff}$ = 3.87 $\mu_{\rm B}$, is close to
the value g$\sqrt{S(S+1)}$ = 3.87 $\mu_{\rm B}$ expected for $S$ = 3/2 moments of free Cr$^{3+}$ ions.
The negative $\theta_{\rm CW}$ indicates the presence of dominant antiferromagnetic interactions between the $S$ = 3/2 spin of Cr$^{3+}$
ions. 
%The given value of core diamagnetic susceptibility
%of the individual ions in ref \cite{Bain2008} were used to calculate
%core diamagnetic susceptibility and we obtained $\chi_{\rm core}$ = $-$ 1.18 $\times$ 10$^{-4}$ cm$^{3}$/mol
%and $\chi_{vv}$ = ($\chi_{0}$ $-$ $\chi_{\rm VV}$) = $-$ 7.62 $\times$ 10$^{-4}$ cm$^{3}$/mol.
 As depicted in the inset of Fig.~\ref{KCTPO2} (a), the Curie-Weiss fit begins to diverge below 80 K, signaling the onset of antiferromagnetic spin correlations upon lowering the temperature \cite{PhysRevB.105.094439}. Interestingly, the material shows anomalies at low temperatures associated with magnetic phase transitions.\\
To further understand the effect of magnetic fields on the long-range magnetic ordered state in KCTPO, the zero-field-cooled (ZFC) and field-cooled (FC) $\chi$ measurements were performed in weak magnetic fields $\mu_{0}H \geq 0.005$ T. Figure~\ref{KCTPO2} (b) reveals the presence of a clear bifurcation between ZFC and FC $\chi$ below the temperature ($T_{\rm L}$) = 4.3 K in a magnetic field of $\mu_{0}H = 0.005$ T,  which implies the presence of a ferromagnetic component below $T_{\rm L}$. Nevertheless, as the magnetic field strength increases, the deviation between ZFC and FC magnetic susceptibility gradually diminishes, almost converging at a magnetic field $\mu_{0}H$ = 0.05 T.  \\
In order to shed more insights concerning the nature of magnetically ordered phases at different temperature  regimes, isotherm magnetization measurements were performed at several temperatures as shown in the inset of Fig.~\ref{KCTPO2} (c). Several features can be observed in the isotherm magnetization curves such as a weak $S$-shaped curvature in low-fields at 2 K, indicating the existence of a ferromagnetic component consistent with the bifurcation between ZFC and FC susceptibility below $T_{\rm L}$ \cite{Zhu2021,Bukowski2022,PhysRevB.103.184429}. Notably, the magnetization at 2 K does not reach saturation in a high magnetic field (9 T); instead, the linear behavior of the magnetization curves implies the presence of a canted  antiferromagnetic state below $T_{\rm L}$. A similar scenario, attributed to the canted antiferromagnetic phase, has also been observed in several frustrated magnets  \cite{PhysRevB.87.155136,PhysRevB.101.054425}.
As the temperature increases, the $S$-shaped curvature is gradually suppressed due to the quenching of the ferromagnetic component, finally disappearing. The resulting magnetization curve becomes linear at 10 K, indicating the crossover to an additional antiferromagnetic phase. These observations imply that KCTPO exhibits a canted antiferromagnetic state below $T_{\rm L}$,  above which it  hosts an additional antiferromagnetic state corroborated by the specific heat results described below. To find the saturation magnetic moment in high magnetic fields, magnetization measurements were performed at 4 K using a pulsed field magnet up to 30 T.\\ Figure~\ref{KCTPO2} (c) represents the high-field magnetization data at 4 K, calibrated using the VSM-SQUID data  at 4 K. It is worth to note that a weak ferromagnetic component persists in low fields $\mu_{0}H$ $\leq$ 0.012 T and beyond this magnetic-field, the magnetization  continues to increase linearly  until $\mu_{0}H$ $\leq$ 14 T and reaches a saturation magnetic moment of 3.04 $\mu_{\rm B}$
in 24 T. We recall that the saturation field is comparable to the CW temperature discussed above. The
obtained saturation magnetic moment of 3.04 $\mu_{\rm B}$ is close to the expected
theoretical value of 3.0 $\mu_{\rm B}$ per Cr$^{3+}$ ions ($S$ = 3/2). The absence of a fractional ``1/3" magnetization plateau indicates either the classical behavior of Cr$^{3+}$ spins or the presence of perturbation terms such as magnetic anisotropy and longer-range interactions \cite{SCHMIDT20171,PhysRevB.105.L180405}.

\begin{figure*}
	\centering
	\includegraphics[width=\textwidth]{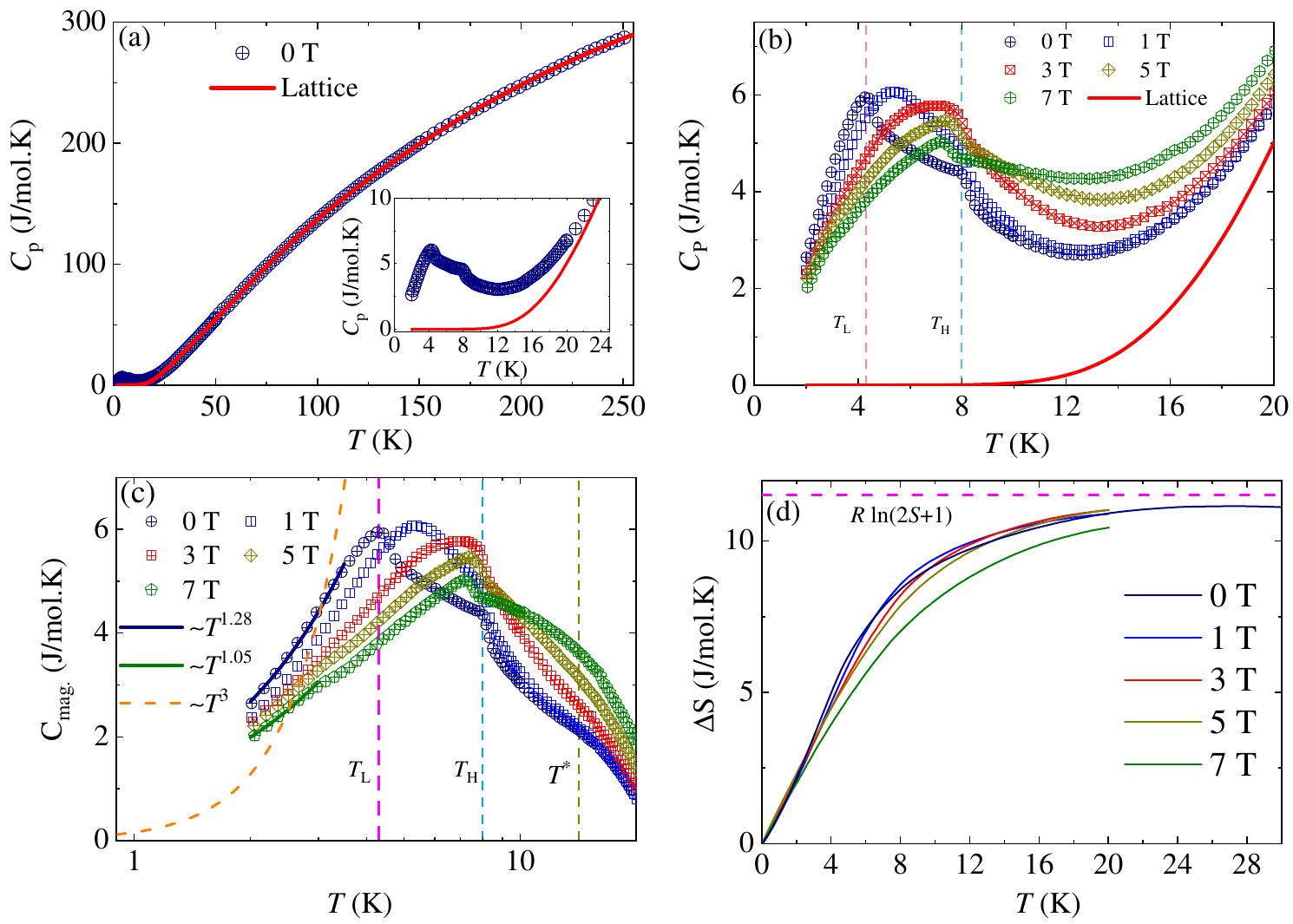}
	\caption{(a)Temperature dependence of specific heat (\textit{C}$_{p}$) of  K$_{2}$CrTi(PO$_{4}$)$_{3}$  in zero-magnetic field. The solid red line represents the lattice contributions obtained by combining one Debye and three Einstein functions as described in text. The inset shows a closer view of the low-temperature specific heat. (b) 
		Temperature dependence of $C_{p}(T)$ in several magnetic fields in the temperature range 2 K $\leq$ $T$ $\leq$ 20 K. The dashed vertical lines indicate two anomalies: one at temperature $T_{\rm L}$ = 4.30 K and another at temperature $T_{\rm H}$ = 8 K in zero-field.   (c) Temperature dependence of $C_{\rm mag}(T)$ in several magnetic fields in semi-log scale. Below $T_{\rm L}$ = 4.3 K,  the solid lines tentatively represent  $\sim T^{n}$  power-law behavior of magnetic specific heat, where $n$ varies from 1.3 to 1 in an applied  magnetic field of 7 T. The dashed orange line represents the $C_{\rm mag}(T)$ $\sim$ $T^{3}$
		behavior typical for conventional antiferromagnets.  (d) Temperature dependence of entropy change in several magnetic fields with the horizontal pink line indicating the expected entropy of $R$ln(4) for $S$ = 3/2 spin of Cr$^{3+}$ ions. 
	 }{\label{KCTPO3}}.
\end{figure*} 
\subsection{Specific heat}
Specific heat is an excellent probe to track magnetic order and associated low-lying excitations in frustrated magnets.
In order to discern the  evidence of magnetic long-range order, temperature-dependent specific heat measurements were performed in several magnetic fields. Figure~\ref{KCTPO3} (a)  shows the temperature dependence of  specific heat in the whole measured temperature range in zero-magnetic field. As shown in Fig.~\ref{KCTPO3} (b),  below $T$ = 10 K, the specific heat exhibits two anomalies: one at the temperature $T_{\rm L}$ = 4.3 K and another at the temperature $T_{\rm H}$ = 8 K in zero-magnetic field,  suggesting the occurrence of successive magnetic phase transitions in KCTPO. 
It is worth to note that the low-temperature anomaly occurring at $T_{L}$  in specific heat coincides with the temperature below which the splitting of ZFC and FC $\chi$ (Fig.~\ref{KCTPO2} (b)) as well as the appearance of a ferromagnetic component in magnetization are observed (Fig.~\ref{KCTPO2} (c)). This coincidence suggests that the anomaly at $T_{L}$ is most likely related to the transition temperature of the canted antiferromagnetic phase in KCTPO. On the other hand, the anomaly at $T_{\rm H}$ = 8 K is attributed to the transition temperature of additional antiferromagnetic phase of KCTPO. We recall that the two successive magnetic phase transitions are also observed in the two coupled trillium lattice K$_{2}$Ni$_{2}$(SO$_{4}$)$_{3}$ \cite{PhysRevLett.127.157204}.    \\
Figure~\ref{KCTPO3} (b) displays the specific heat in several magnetic fields in the low-temperature regime. The salient observation is that the low-temperature anomaly at $T_{\rm L}$ shifts towards higher temperatures upon the application of a magnetic field, eventually vanishing at a magnetic field of $\mu_{0}H$ = 3 T. The increase in $T_{\rm L}$  with increasing magnetic field can be ascribed to either strong quantum fluctuations or the presence of competing anisotropic interactions.   While the applied magnetic field increases, the anomaly at $T_{\rm H}$ widens, shifts to slightly lower temperatures, and a clear $\lambda$-like anomaly is observed in $\mu_{0}H$ = 7 T, suggesting the presence of a typical antiferromagnetic  second-order phase transition at $T_{\rm H}$ in KCTPO. Moreover, in the temperature range 8 K $\leq$  $T$  $\leq$  20 K, the specific heat exhibits field dependency, indicating that specific heat is of magnetic origin in this temperature range. \\
In order to estimate the magnetic entropy and understand the  low-temperature magnetic properties relevant to this trillium lattice antiferromagnet, one must subtract the lattice specific heat ($C_{\rm latt}(T)$) from the total specific heat ($C_{\rm p}(T)$). To extract the lattice contributions, we used a combination of one Debye  and three Einstein terms, i.e., \begin{equation*}
C_{\rm latt}(T)=C_{D}[9k_{B} \left(\frac{T}{\theta_{D}}\right)^{3}\int_{0}^{\theta_{D}/T}\frac{x^{4}e^{x}}{(e^{x}-1)^{2}}dx]
\end{equation*}
\begin{equation}\label{eqn:debye}
+\sum_{i=1}^{3} C_{E_{i}}[3R\left(\frac{\theta_{E_i}}{T}\right)^{2}\frac{\text{exp}(\frac{\theta_{E_{i}}}{T})}{(\text{exp}(\frac{\theta_{E_{i}}}{T})-1)^{2}}], 
\end{equation}
where $\theta_{D}$ is the Debye temperature, $\theta_{E_is}$ are the Einstein temperatures of the three modes, \textit{R} and \textit{k}$_{B}$ are the molar gas constant and Boltzmann constant, respectively. The corresponding lattice fit, yielding $\theta_{D}$ = 223 K, $\theta_{E_1}$ = 376 K, $\theta_{E_2}$ = 615 K, and $\theta_{E_3}$ = 1344 K, is represented by the solid red line on top of the experimental data point in Fig.~\ref{KCTPO3} (a). 
In order to minimize the fitting parameters, the coefficients were set at a fixed ratio of $C_{\rm D}$ : $C_{E_{1}}$ : $C_{E_{2}}$ : $C_{E_{3}}$ = 1 : 1 : 1.5 : 6, closely matching with the ratio of the number of heavy atoms (K, Cr, Ti, P) to light atoms (O) in KCTPO.\\
\begin{figure*}
	\centering
	\includegraphics[width=\textwidth]{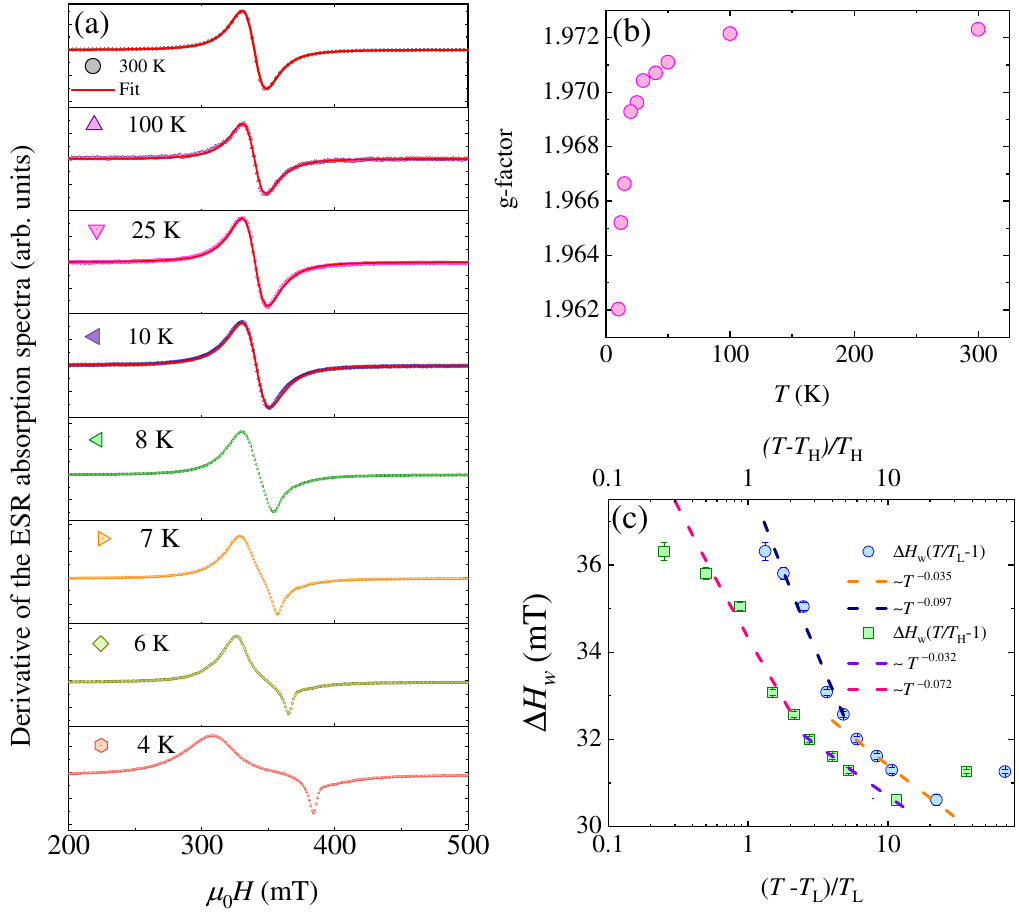}
	\caption{(a) Derivative of the ESR absorption spectra of K$_{2}$CrTi(PO$_{4}$)$_{3}$ at selected temperatures. The solid red lines indicate the fitting to the Lorentzian line shape above 8 K.  (b) Temperature dependence of g-factor in semi-log scale.   (c)  Semi-log plot of the ESR linewidth $\Delta H_{\rm w}$ as a function of the reduced
		temperature expressed as $T_{\rm rl}$ = ($T-$ $T_{\rm L}$)/$T_{\rm L}$ on the bottom $x$-axis and $T_{\rm rh}$ = $(T-T_{\rm H})/T_{\rm H}$ on the upper $x$-axis. The dashed lines are power-law fits.
	}{\label{KCTPO4}}.
\end{figure*} 
After subtraction of the lattice contributions, the obtained temperature dependence of magnetic specific heat $C_{\rm mag}$ ($T$) is shown in Fig.~\ref{KCTPO3} (c).
As the temperature decreases in a zero-magnetic field, $C_{\rm mag}$ increases and  exhibits a broad maximum around temperature $T^{*}$ = 14 K, followed by a weak kink at $T_{\rm H}$. The broad maximum is attributed to the presence of short-range spin correlations \cite{PhysRevLett.127.157204}. 
The presence of a moderate frustration parameter $f$, roughly estimated through the ratio $f = |\theta_{\rm CW}|/T_{\rm H} \approx 3$, could account for both the short-range order and the suppression of magnetic order. The anomaly at $T_{\rm H}$ is relatively weak compared to the anomaly at $T_{\rm L}$ similar to the observed anomalies in K$_{2}$Ni$_{2}$(So$_{4}$)$_{3}$ \cite{PhysRevLett.127.157204}. Nevertheless, upon increasing the magnetic field, the canted antiferromagnetic phase diminishes, leading to an increase in the anomaly observed at $T_{\rm H}$ in $\mu_{0}H$ = 7 T.
Most notably, as shown in Fig.~\ref{KCTPO3} (c), the magnetic specific heat follows $C_{\rm mag}~$$\sim$ $T^{n}$ ($n$ = 1.28) power-law in zero field for temperatures below $T_{\rm L}$, which significantly deviates from the $C_{\rm mag}(T)$ $\sim$ $T^{3}$ behavior typical of conventional antiferromagnets. In the presence of a magnetic field up to 7 T, the average value of $n$ was found to be 1.10. It is worth noting that in future low temperature  specific heat measurements are required to confirm whether the power-law behavior persists even in the sub-Kelvin temperature range. In contrast to KCTPO, the trillium lattice structure  with two interpenetrating sublattices of K$_{2}$Ni$_{2}$(SO$_{4}$)$_{3}$, well below the transition temperature, the magnetic specific heat follows $C_{\rm mag}$ $\sim$ $T^{2}$ behavior \cite{PhysRevLett.127.157204}.  \\
\begin{figure*}
	\centering
	\includegraphics[width=\textwidth]{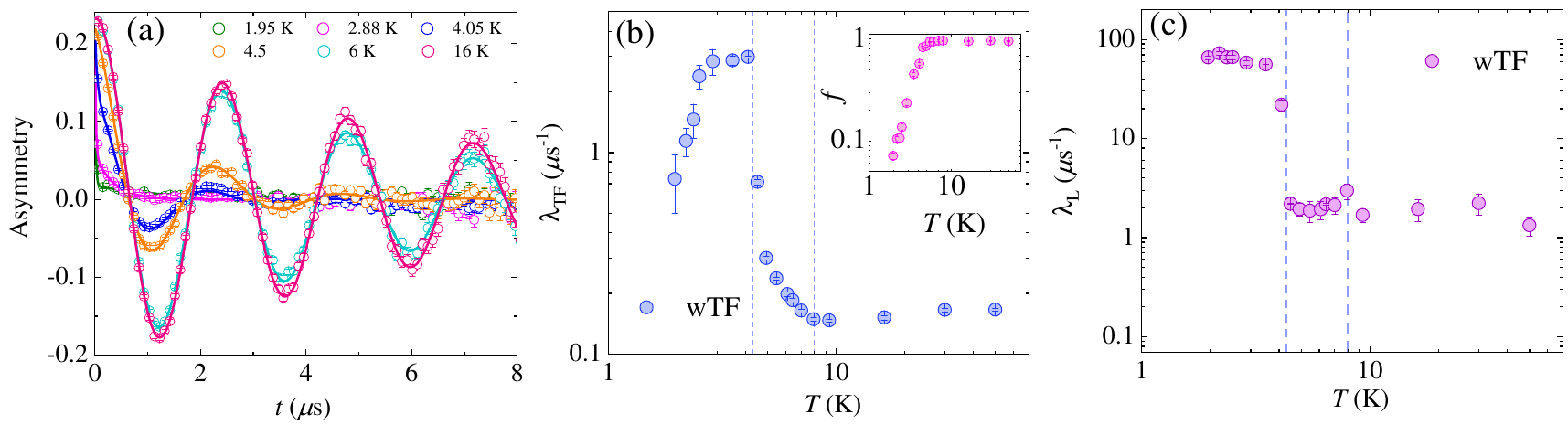}
	\caption{(a) Time dependence of the muon asymmetry in K$_{2}$CrTi(PO$_{4}$)$_{3}$ in a weak transverse field ($B_{\rm TF}$ = 32 G) at a several temperatures. The solid lines are fits
		obtained using Eq.~\ref{WTF}.  (b) Temperature dependence of  muon spin relaxation rate in transverse applied field on a double logarithmic scale. The inset shows the fraction of the oscillating component as a function of temperature. (c) Temperature dependence of  muon spin relation rate due to internal magnetic field in the investigated sample. The dashed vertical lines indicate  the position of characteristic transition temperatures at $T_{\rm L}$ = 4.3 K and $T_{\rm H}$ = 8 K.    
	}{\label{KCTPO5}}.
\end{figure*}
To quantify the entropy release associated with phase transitions and spin dynamics, it is crucial to account the magnetic-specific heat below 2 K. In the absence of no more phase transition below 2 K, we extend the corresponding power-law to zero temperature.  Next, the entropy was obtained by integrating $C_{\rm mag}/T$ with respect to temperature, as shown in Fig.~\ref{KCTPO3} (d). The obtained saturation entropy is found to be 11.16 J/mol$\cdot$K at 30 K that is close the theoretically expected value 11.52 J/mol$\cdot$K for $S$ = 3/2  moments.  The missing  $\sim$ 3.12 \text{\%} entropy  is ascribed to either the existence of short-range magnetic correlations well above the transition temperature or an overestimation of the lattice contribution given the limitations of the model employed here \cite{Khatua2021,PhysRevB.107.214411}.  Additionally, around the transition temperature $T_{H}$ of 8 K, approximately 76 \text{\%} of the saturation entropy is released, suggesting that roughly 20 \text{\%} originates from short-range spin correlations that begin forming as high as at $\sim 3T_{\rm H}$. However, the lack of a few percentage of entropy because of short-range order persisting above the transition temperature is typical for frustrated magnets \cite{PhysRevB.105.094439,Khatua2021}. This observation is corroborated by the ESR results presented in the following subsection.
\subsection{Electron spin resonance}
To provide  microscopic insights into the temperature evolution of spin correlations, we conducted X-band ESR measurements down to 4 K \cite{abragam2012electron}. The obtained ESR spectra at several temperatures are shown in Fig.~\ref{KCTPO4} (a), which fits well with a  derivative of a Lorentzian curve above 8 K. This indicates that the ESR signal is exchange narrowed.
 At temperatures below $T$ = 10 K, the spectra cannot be fitted by a single
 Lorentzian curve, indicating the development of antiferromagnetic resonance mode and supporting the conclusions drawn from thermodynamic results. The temperature dependence of estimated $g$-factors from the fit is shown in Fig.~\ref{KCTPO4} (b). Above $T$ = 100 K, the obtained $g$ value remains relatively constant at approximately $g$ = 1.972, slightly lower than the standard $g$ = 2 for a free Cr$^{3+}$ ion. This value is typical for a less-than-half-filled Cr$^{3+}$ ion with a negligible spin-orbit interaction in an octahedral ligand coordination, as observed in similar compounds \cite{PhysRevB.107.214411}. The $g$-factor starts to decrease with temperature below 100 K, which corresponds to the deviation from the CW fit (see the inset of Fig.~\ref{KCTPO2} (a)). This is attributed to the persistent magnetic correlations up to several $|\theta_{\rm CW}|$, often observed in frustrated magnets \cite{PhysRevB.105.094439}.\\
The ESR linewidth is directly proportional to the spin correlations in transition metal-based systems with minimum spin-orbit coupling.  Figure \ref{KCTPO4} (c) shows the estimated ESR linewidth ($\Delta H_{\rm w}$) of KCTPO as a function of  reduced temperatures  $T_{rl}$ = $(T-T_{\rm L})/T_{\rm L}$ in the bottom $x$-axis and  $T_{rh}$ = $(T-T_{\rm H})/T_{\rm H}$  in the upper $x$-axis. Above the transition temperature $T_{\rm H}$, the observed linewidth exhibits a power-law behavior of $T_{\rm rl}/ T_{\rm rh} \sim T^{-p}$ in two distinct temperature ranges. For temperatures spanning from 10 K to 24 K, the power exponent $p$ takes values of  0.097 and 0.072, while in the range of 30 K to 100 K, the values of $p$ are 0.035 and 0.032 for $T_{\rm rl}$ and $T_{\rm rh}$, respectively \cite{PhysRevB.95.184430,PhysRevB.107.214411,PhysRevB.93.174402,PhysRevB.105.094439}. The critical-like broadening in the temperature range 10 K $\leq$ $T$ $\leq$ 24 K, alongside the weak hump at $T^{*}$ in $C_{\rm mag}$ and the small percentage of entropy released above 8 K, supports the notion of the critical-like spin fluctuations.  Conversely, the broadening observed in the temperature range of 30 K $\leq$ $T$ $\leq$ 100 K is ascribed to cooperative paramagnetic behavior between the $S=$ 3/2 spin of Cr$^{3+}$ ions. Additionally, a deviation from the Curie-Weiss fit has also been noted below 100 K (refer to Fig.~\ref{KCTPO2} (a)). The presence of two stage broadening is also observed in other high-spin based frustrated magnets alluding to the two-step thermal evolution of magnetic correlations \cite{PhysRevB.95.184430,PhysRevB.105.094439,PhysRevB.93.174402}. 
\subsection{Muon spin relaxation}
To capture the microscopic details of spin dynamics and the internal magnetic field distribution, $\mu$SR measurements were performed on polycrystalline samples of KCTPO in zero-field (ZF) and weak transverse field (wTF). When a weak transverse field (here $B_{\rm TF}$ = 32 G) is applied perpendicular to the initial direction of the muon spin polarization, the implanted muon spin precesses around the applied transverse field in the paramagnetic state of the sample, leading to an oscillatory signal  with a frequency of $\gamma_{\mu}$$B_{\rm TF}$/2$\pi$, where $\gamma_{\mu}$ = 2$\pi$ $\times$ 135.5 MHz/T represents the muon gyromagnetic ratio \cite{yaouanc2011muon}. In addition, when a static magnetic field is present in the sample, the muon spin polarization exhibits a non-oscillatory signal, as observed in the material under investigation \cite{Nocerino2023,Khatua2021,PhysRevB.106.214410}. Figure~\ref{KCTPO5} (a) shows the evolution of the TF spectra at selected temperatures that were fitted according to 
\begin{equation}\label{WTF}
A_{\rm wTF} (t) = A_{0}f \text{cos} (\gamma_{\mu}B_{\rm TF}t + \phi) e^{-\lambda_{\rm TF}t} + (1-f) A_{0}e^{-\lambda_{\rm L}t},
\end{equation}
which combines an exponentially decaying oscillatory component, corresponding to muon spins experiencing zero-static field, and an exponentially decaying non-oscillatory component that accounts for the component of the static local field  parallel to the initial muon polarization. In Eq.~\ref{WTF}, $A_{0}$ represents the initial asymmetry at time zero, $\phi$ denotes the relative phase, $\lambda_{\rm TF}$ stands for the muon spin relaxation rate in the applied transverse field, and $f$ quantifies the fraction of the oscillatory component while  $\lambda_{\rm L}$ refers to the muon spin relaxation rate caused by the internal magnetic fields. At short time scales, we observe a significantly damped signal with decreasing temperature without any loss of initial asymmetry, which is a characteristic signature of a magnetically ordered state in KCTPO. \\
The estimated fraction of the oscillatory signal $f$, associated to the volume fraction experiencing zero-static field (inset of Fig.~\ref{KCTPO5} (b)), remains constant to  a value of  0.93 in the temperature range 4.3 K $\leq$  $T$ $\leq$ 78 K. This temperature independent value of $f<1$  is attributed to the presence of weak static local fields above 4.3 K which is also observed in ESR results.   However, it gradually decreases below $T_{\rm L}$= 4.3 K, indicating a crossover to anther magnetically ordered state. It is noteworthy that $f$ does not sharply drop to zero below the transition temperature $T_{\rm L}$, suggesting the existence of dynamic local fields in the ordered state. Upon lowering the temperature, the transverse field muon spin relaxation  rate gradually increases (see Fig.~\ref{KCTPO5} (b)), as expected in the extreme motional narrowing limit of fluctuating magnetic moments. A sharp peak in $\lambda_{\rm TF}$ at $T_{\rm L}$ indicates the critical slowing down of spin fluctuations above the transition temperature as evident from unsaturated magnetic entropy. The notable observation is the occurrence of the critical spin relaxation below $T_{\rm H}$ \cite{PhysRevB.104.045145}. Conversely, the muon spin relaxation rate caused by the static local field (Fig.~\ref{KCTPO5} (c)) exhibits a weak kink at $T_{\rm H}$, attributed to the antiferromagnetic phase transition similar to that observed in specific heat. This is followed by a rapid rise towards $T_{\rm L}$, indicating the enhancement of  internal field fluctuations through spin reconfiguration to the canted antiferromagnetic state. It is interesting to note that, below $T_{\rm L}$, the $\lambda_{\rm TF}$ remains constant down to 1.93 K without displaying a sharp fall, persistent spin fluctuations even in the magnetically ordered state, similar to that observed in coupled trillium lattice K$_{2}$Ni$_{2}$(SO$_{4}$)$_{3}$ \cite{PhysRevLett.127.157204}. \\
\begin{figure*}
	\centering
	\includegraphics[width=\textwidth]{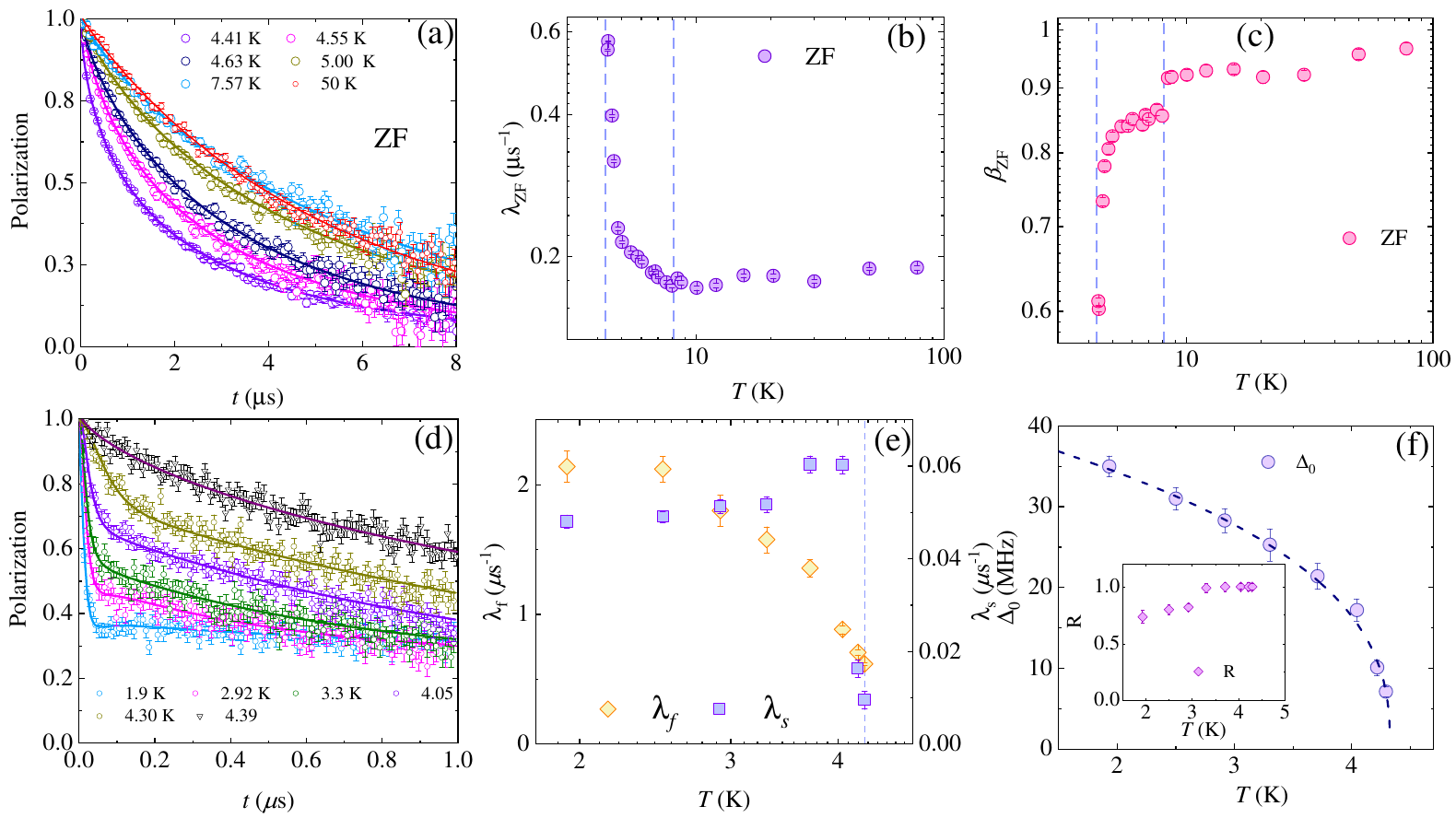}
	\caption{(a) Time dependence of zero-field $\mu$SR spectra of K$_{2}$CrTi(PO$_{4}$)$_{3}$ at temperatures higher than the temperature of canted antiferromagnetic phase transition. The solid lines represent the fitted curve by a stretched exponential function. (b), (c)  Temperature dependence of  muon spin relaxation rate and stretched exponent as a function of 
		temperature in zero magnetic field on a double logarithmic scale. The dashed vertical bars indicate  the position of characteristic transition temperatures at $T_{\rm L}$ = 4.30 K and $T_{\rm H}$ = 8 K.     (d) Time evolution of zero-field spectra at temperatures below the canted antiferromagnet state. The solid lines are the fits of Eq.~\ref{bdg} to the data. (e) Temperature dependence of fast (left y-axis) and slow (right y-axis) muon spin relaxation rate  on a double logarithmic scale. (f)  Mean value of the Gaussian-broadened Gaussian distribution as a function of  temperature where the solid lines are the fitted curves with critical scaling behavior of internal field distribution  with critical exponent $\beta$ = 0.37. Inset shows the temperature dependence of  $R$ i.e, the ratio
		of the width of distributions (W) to the distribution mean($\Delta_{0}$).
	}{\label{KCTPO6}}.
\end{figure*} 
To validate the spin dynamics in the ordered states, ZF-$\mu$SR measurements were performed in the wide temperature range.   
 Figure~\ref{KCTPO6} (d) shows the ZF-$\mu$SR spectra at temperatures above $T_{\rm L}$ without any change of the initial asymmetry.
 Furthermore, despite an anomaly in specific heat associated to the magnetic phase transition,  the absence of any oscillation or a strong damped signal even in short-time scale, suggests that a dynamical behavior of electronic spins associated with the magnetic phase similar to that observed in trillium lattice antiferromagnet K$_{2}$Ni$_{2}$(SO$_{4}$)$_{3}$.  The corresponding $\mu$SR data (Fig.~\ref{KCTPO6}) can be well fitted to the phenomenological model $P(t)$ = $A_{0}$ e$^{-(\lambda_{\rm ZF} t)^{\beta_{\rm ZF}}}$, where $A_{0}$ is the initial asymmetry, $\lambda_{\rm ZF}$ is the muon spin relaxation rate in zero-field and $\beta_{\rm ZF}$ is the  stretched
 exponent. The estimated relaxation rate  exhibits several features as shown in Fig.~\ref{KCTPO6} (b).  Upon lowering the temperature from 78 K, the $\lambda_{\rm ZF}$ remains constant  in the temperature range 8 K $\leq$ $T$ $\leq$ 78 K. The essentially same behavior is observed by $\lambda_{\rm TF}$ (see Fig.~\ref{KCTPO5} (b)). Below 8 K, the $\lambda_{\rm ZF}$ shows a rapid increase, ascribed to the 
presence of the canted antiferromagnetic phase transition accompanied by a critical slowing down of spin fluctuation consistent with our wTF-$\mu$SR results. The obtained stretched exponent, measuring the distribution of electronic moment, is shown Fig.~\ref{KCTPO6} (c). At high-temperatures, $\beta_{\rm ZF}$ is close to one, corresponding to fast fluctuation limit. As the temperature decreases, the $\beta_{\rm ZF}$ value gradually decreases, notably dropping to 0.85 at $T_{\rm H}\sim$ 8 K.  Below 8 K, the $\beta_{\rm ZF}$ value gradually decreases, but there is a sudden drop at $T_{\rm L}$ to a value of 0.6.   
It is observed that the ZF-$\mu$SR data (for $T\geq 4.41$ K) cannot be accurately fitted using a single exponential function ($\beta_{\rm ZF}$ = 1) or models with multi-exponential components \cite{PhysRevB.105.094439}.  The obtained $\beta_{\rm ZF} < 1$ is associated with the presence of a distribution of relaxation times, as expected when various spins fluctuate in different timescales due to competing magnetic interactions in KCTPO. The observed $\beta_{\rm ZF}$ of 0.6  ($>1/3$) value, as seen in Fig.~\ref{KCTPO6} (c), implies that the splitting in ZFC and FC $\chi$ data at 4.3 K is due to canted antiferromagnetic phase rather than spin freezing \cite{PhysRevLett.72.1291}. 
\\
Figure~\ref{KCTPO6} (d) displays the time evolution of ZF-$\mu$SR spectra at several temperatures below 4.5 K in the short-time scale. As the temperature decreases below 4.5 K, the polarization initially exhibits a moderate departure from stretched exponential behavior \cite{PhysRevB.77.092403}. As it reaches around 1.9 K, it undergoes an abrupt transition into a strongly non-exponential signal.
The shorter time domain clearly shows a substantially damped signal, making it challenging to simulate the muon polarization throughout the whole temperature range using a stretched exponential relaxation function. Furthermore, the absence of any peak in the first Fourier transform of the corresponding spectra indicates that the damped signal can not be attributable to coherent oscillations typically observed in long-range ordered magnets \cite{PhysRevB.97.224416,weinhold2023magnetism,PhysRevB.106.214410}. Instead, it suggests the existence of disordered static moments of Cr$^{3+}$ ions below 4.3 K. After attempting with several models \cite{Li2021,PhysRevB.96.094432,PhysRevLett.104.057202}, it turns out that the ZF-$\mu$SR data can be modeled with  the
following polarization function 
\begin{equation}\label{bdg}
P_{\rm ZF}= f_{\rm z}P_{\rm GDG}(t; \Delta_{0}, W)e^{-\lambda_{s}t} +(1-f_{\rm z})e^{-\lambda_{\rm f}t},
\end{equation}
which takes into account a phenomenological Gaussian-broadened Gaussian (GBG) function with a slow exponential decay with relaxation rate $\lambda_{\rm s}$ and a fast exponential decay with relaxation rate $\lambda_{\rm f}$ \cite{PhysRevB.56.2352,PhysRevB.98.014407}. The parameter $f_{\rm z}$ defines the fraction of Cr$^{3+}$ moments that produce static local fields, while the remaining part contributes to dynamic fields. The following GBG polarization function
\begin{equation}
\begin{split}
P_{\rm GBG}(t) =& \frac{1}{3}+\frac{2}{3}(\frac{1}{1+R^{2}\Delta_{0}^{2}t^{2}})^{3/2}(1-\frac{\Delta_{0}^{2}t^{2}}{1+R^{2}\Delta_{0}^{2}t^{2}}) \\
&\rm exp [-\frac{\Delta_{0}^{2}t^{2}}{2(1+R^{2}\Delta_{0}^{2}t^{2})}]\\
\end{split}
\end{equation} 
 is an extension of the Gaussian Kubo-Toyabe function with a Gaussian distribution width $W$ and a mean value $\Delta_{0}$, where $R$ = $W/\Delta_{0}$. This function is  commonly  employed when the internal field distribution exceeds the Gaussian field distribution. It effectively captures the existence of disordered static magnetic moments with short-range correlations as observed in 3D hyperkagome compound such as Na$_{4}$Ir$_{3}$O$_{8}$ which exhibits ZFC and FC splitting in dc susceptibility similar to the compound studied here \cite{PhysRevLett.113.247601}. The extracted temperature dependence of fast relaxation rate ($\lambda_{\rm f};$ left $y$-axis), representing a few fraction of Cr$^{3+}$ moments exhibiting dynamic behavior, and the slow relaxation rate ($\lambda_{s};$ right $y$-axis) corresponds to the remaining fraction Cr$^{3+}$ moments displaying static behavior are shown in Fig.~\ref{KCTPO6} (e). Both $\lambda_{\rm f}$ and $\lambda_{s}$ start to increase below $T_{\rm L}$ and attain a constant value at low temperatures indicating the coexistence of static and dynamic components of Cr$^{3+}$ moments in KCTPO.\\ In KCTPO, the origin of static moments can be associated to the  canted antiferomagnetic phase involving a small fraction of static Cr$^{3+}$ moments. Simultaneously, the rest of the Cr$^{3+}$ moments maintain a dynamic state most likely due to the interplay of competing magnetic interactions. In geometrically frustrated magnets, the coexistence of dynamic and static electronic moments is a common scenario, as also observed in two coupled trillium lattice K$_{2}$Ni$_{2}$(SO$_{4}$)$_{3}$ and pyrochlore lattice NaCaNi$_{2}$F$_{7}$ \cite{Cai_2018}. 
% It
% should be mentioned that the GBG polarization function effectively describes the spin dynamics of quantum magnets featuring inhomogeneous static magnetic moments owing to the random distribution of magnetic sites \cite{Lee2023,PhysRevB.98.014407}. This suggests a potential of a random distribution of magnetic sites in KCTPO, although our XRD results rules out a scenario. This calls for future neutron diffraction experiments to further investigate this discrepancy. 
 $\Delta_0$ undergoes an order-parameter-like decrease as the temperature increases from 1.9 K towards $T_{\rm L}$.  It may be noted that the relative distribution width $R$ steadily increases and reaches a value of one at $T_{\rm L}$ (inset of Fig.~\ref{KCTPO6} (f)). The critical-like behavior of the local field distribution near $T_{\rm L}$ is described by phenomenological expression \cite{Lee2023,PhysRevB.84.174403} $\Delta_{0}(T) = \Delta_{0}(T= 0 \ \ \text{K}) \left[1-\frac{T}{T_{\rm L}}\right]^{\beta}$ that yields $\beta$ = 0.37 that is typical for 3D Heisenberg magnets (see the dashed line in Fig.~\ref{KCTPO6} (f)). Using $\Delta_{0}$ = 35 MHz at $T$ = 1.9 K, the calculated internal magnetic field is found to be  $\langle B_{\rm loc}\rangle$ $\approx$ 400 G. In order to confirm such static magnetic field is of electronic origin, we also performed $\mu$SR measurements at 3.3 K in longitudinal field (not shown here), which decouples the static field owing to nuclear origin. It is observed that approximately 2 kG $>$ $\langle B_{\rm loc}\rangle$  longitudinal field is required 
 to recover full polarization, confirming the presence of both dynamic and static moments of Cr$^{3+}$ ions in KCTPO. Similar to KCTPO, the observed muon spin asymmetry of triangular lattice antiferromagnet NiGa$_{2}$S$_{4}$ was characterized by a stretched exponential function at high temperatures, associated to multichannel relaxation, while the data at low temperatures were primarily influenced by static local fields \cite{PhysRevB.77.092403}.
  The contrasting time evolution of the muon spin polarization function, observed both above and below the characteristic temperature of $T_{\rm L}$, may be associated with distinct magnetic phases in KCTPO, similar to those found in other 3D frustrated magnets \cite{PhysRevB.102.144409,PhysRevLett.88.077204,Lee2021,PhysRevMaterials.4.074405}.\\
\section{Conclusion}
In summary, we have thoroughly investigated the structural and magnetic properties of  nearly
perfect $S$ = 3/2 trillium lattice K$_{2}$CrTi(PO$_{4}$)$_{3}$ through magnetization,
specific heat, as well as ESR and $\mu$SR techniques. 
 Magnetic susceptibility  results suggest that the presence of antiferromagnetic interactions between $S$ = 3/2 moments of Cr$^{3+}$ ions.  Specific heat measurements unveil two consecutive phase transitions, one at $T_{\rm L}$ = 4.3 K, corresponding to the canted antiferromagnetic phase, and another at $T_{\rm H}$ = 8 K due to an antiferromagnetic phase. Above $T_{\rm H}$, the  existence  of short-range magnetic order is supported by the observation of maximum entropy release and the manifestation of critical-like behavior in the ESR line width.  Furthermore, $\mu$SR provides concrete evidence of the presence of static and dynamic local magnetic fields, particularly below $T_{\rm L}$, as well as a critical slowing down of spin dynamics above $T_{\rm L}$ and persistent spin dynamics to the lowest
 temperatures. Further investigations on single crystals is essential to gain a detailed understanding of the low-energy excitations associated to the short range order above the transition temperature and the multi-stage spin dynamics in this three-dimensional frustrated antiferromagnet.
    \section*{Acknowledgments}
     P.K. acknowledges the funding by the Science and
    Engineering Research Board, and Department of Science and Technology, India through Research Grants. The work at SKKU was supported by the National Research Foundation
    (NRF) of Korea (Grant no. RS-2023-00209121, 2020R1A5A1016518). We further acknowledge the support of HLD at HZDR, member of the European Magnetic Field Laboratory (EMFL).
\bibliographystyle{apsrev4-1}
\bibliography{KCTPO}
\end{document}